\documentclass[twocolumn,showpacs,preprintnumbers,amssymb]{revtex4}
\usepackage{graphicx}
\usepackage{dcolumn}
\usepackage{bm}

\begin{document}
\draft          
\title{Scarred Resonances and Steady Probability Distribution in a Chaotic Microcavity}
\author{Soo-Young Lee$^1$}
\email{sooyoung@phys.paichai.ac.kr}
\author{Jung-Wan Ryu$^{1,2}$}
\author{Tae-Yoon Kwon$^{1,2}$}
\author{Sunghwan Rim$^1$}
\author{Chil-Min Kim$^1$}
\affiliation{$^1$ National Creative Research Initiative Center for Controlling Optical Chaos,\\
Pai-Chai University, Daejeon 302-735, Korea}
\affiliation{$^2$ Department of Physics, Sogang University, Seoul 121-742, Korea}

\begin{abstract}
We investigate scarred resonances of a stadium-shaped chaotic microcavity.
It is shown that two components with different chirality of the scarring pattern
are slightly rotated in opposite ways from the underlying unstable periodic orbit, 
when the incident angles of the scarring pattern are close to
the critical angle for total internal reflection.
In addition, the correspondence of emission pattern with the scarring pattern
disappears when the incident angles are much larger than the critical angle.
The steady probability distribution gives a consistent explanation
about these interesting phenomena and makes it possible to expect the
emission pattern in the latter case. 
\end{abstract}


\maketitle

\narrowtext

Directional lasing emission from a microcavity
has been intensively studied in the last decade 
due to its direct application to optical communications 
and optoelectronic circuits\cite{book1}.  The first realization of
the directional emission has been achieved in the microcavities which
are slightly deformed from circle or sphere\cite{no97}.
In the slightly deformed microcavities it is known
that the light is emitted from a boundary point of the highest
curvature and the direction is tangential to the boundary. 
In ray dynamical viewpoint, this is a result of 
tunneling of the rays confined in the Kolmogorov-Arnold-Moser tori
through the lowest dynamical barrier.

As the microcavity is strongly deformed, the ray dynamics inside
becomes chaotic and lasing modes from
the chaotic cavity can be expected to have complicated patterns
without any directionality due to the complexity of 
the corresponding ray trajectories.
However, the openness of the microcavity enhances
the scarring phenomenon\cite{He84}, localization along unstable periodic orbits,
and the scarred resonance shows a strong directional emission 
from the chaotic microcavity.
The scarred lasing modes are observed in several experimental studies\cite{scar,Re02}, 
and the emission direction from the scarred mode can be 
nontangential due to the Fresnel filtering effect\cite{Re02}.

In a recent study\cite{Lee04}, it was reported that in a spiral-shaped 
chaotic microcavity there are many resonances showing  special localized patterns,
so-called quasiscarred resonances which have, unlike typical scarred resonances, 
no underlying unstable periodic orbit. 
This finding indicates that the openness of the system plays 
a crucial role in the formation of localized pattern, and
implies that the properties of openness should be imprinted in
scarred resonances also.
It is quite interesting and important to uncover the mechanism of
how the openness of the microcavity can change the scarred resonance pattern 
and can determine the emission direction.
This is the motivation of our study.

In this Letter, we illustrate characteristics of the scarred resonances
in a chaotic microcavity of stadium shape and show that 
the scarred resonances can be classified into two categories, i.e., 
type-I and II for convenience,  based on their difference in 
emission property.  
The type-I scarred resonances have a clear correspondence of the emission
pattern with the scarring pattern, e.g., the bouncing point of the
scarring pattern matches well with the emitting point.
However, in the type-II scarred resonances there is no correspondence
between emission and scarring patterns, and the emission pattern is
determined by the structure of steady probability distribution 
(SPD)\cite{Lee04} near the critical line ($p_c=\sin \theta_c=1/n$,
 $n$ being the refractive index) 
for total internal reflection. Moreover, the SPD provides
a consistent explanation about a novel feature of the type-I scarred
resonances, i.e., the spatial splitting phenomenon of  
different chiral components of the scarring pattern.

It is known that long time ray dynamical properties of a chaotic microcavity 
can be described by the SPD, $P_s(s,p)$, which
is the spatial part of the exponentially decaying survival probability 
distribution in phase space $(s,p)$, where $s$ is the boundary coordinate and
 $p=\sin \theta$, $\theta$ being incident angle\cite{Lee04}. In chaotic microcavities
this SPD is a useful tool in identifying both the structure of openness
and the escaping mechanism of ray trajectories. 
In addition, various informations
about the long time ray dynamics can be easily calculated from the SPD, 
e.g., the ray dynamical far field distribution is given by
\begin{equation}
P_{far} (\phi) \propto \int ds dp\, P_s (s,p)\, {\cal T} (p) \,\delta (\phi -f(s,p)),
\label{far}
\end{equation}
where ${\cal T}(p)$ is the transmission coefficient\cite{book2} and the far field angle $\phi$
is given as $f(s,p)$ determined by the geometry of boundary and Snell's law.
This $P_{far} (\phi)$ would correspond to the averaged far field
distribution for the resonances with relatively high quality factor $Q$.
Even in the case of individual resonance pattern,
the SPD provides a background structure in phase space and
is useful to understand the resonance pattern and its emission property.

As a phase space representation of a quantum mechanical eigenfunction,
the Husimi function has been widely used in billiard
systems\cite{Ba04}. For dielectric microcavities,  
Hentschel et al.\cite{He03} have developed the incident and 
the reflective Husimi functions as
\begin{equation}
H^{inc(refl)}(s,p)=\frac{k}{2\pi} \left|
- {\cal F} h(s,p)+(-) \frac{i}{k {\cal F}}h'(s,p) \right |^2,
\label{husimi}
\end{equation} 
where the weighting factor is ${\cal F}= \sqrt{n \sqrt{1-p^2}}$ and
the components of the Husimi function are given as 
$h(s,p) = \int ds' \psi(s')  \xi(s';s,p)$,
$h'(s,p) = \int ds' \partial \psi(s') \xi(s';s,p)$, respectively.
$\psi(s)$ and $\partial \psi(s)$ are the boundary wavefunction and
its normal derivative, and 
$\xi(s';s,p)$ is the minimal-uncertainty wave packet of the form
\begin{equation}
\xi(s';s,p) = \sum_l \frac{1}{\sqrt{\sigma \sqrt{\pi}}}
\exp [-\frac{1}{2\sigma^2}(s'-s)^2-ik p(s'+Ll)]
\end{equation} 
where $k$ and $L$ are the wavenumber inside the cavity and
 the total length of boundary, respectively. This corresponds 
to a Gaussian wavefunction centered at $(s,p)$,
and we set the aspect ratio factor as $\sigma =(\sqrt{2}/k)^{-1/2}$ 
throughout this Letter. 
This formalism for Husimi functions is exact
in the high $k$ limit,and it still gives good approximates around
 $kR \simeq 90$ where our calculation is performed.

\begin{figure}

\vspace{0.3cm}

\hspace{0.cm} \includegraphics[height=5.cm, width=8.2cm]{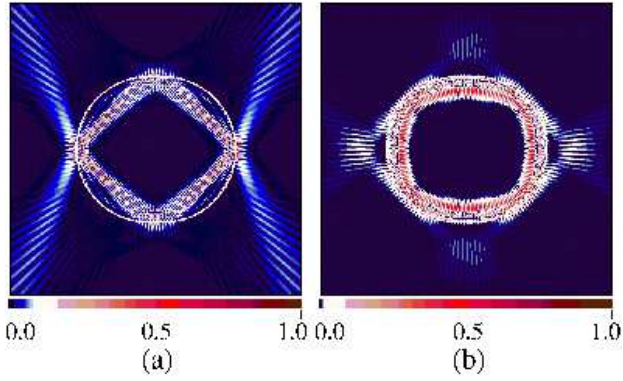}

\vspace{0.0cm}

\caption{(color online) The type-I scarred resonances showing a diamond (a) and 
a rectangle (b) scarring patterns in a stadium-shaped microcavity with
$n=\sqrt{2}$.   Note that the refractive emissions
come out from the bouncing points of the scarring patterns.
The resonance positions are (a) $kR=(91.68,-0.17)$ and (b) $kR=(92.48,-0.035)$. }

\end{figure}

As shown in Fig. 1, by using the boundary element method (BEM)\cite{Wie03}
we obtain two examples of type-I scarred resonances,
where scarring patterns are diamond (a) and rectangle (b),
for a stadium-shaped microcavity with 0.2$R$ length of linear segment,
$R$ being the radius of two semicircles.
It is clear that the emitted light is refracted out 
from the bouncing points of the scarring patterns inside the microcavity.
We also obtain the Husimi functions
 $H^{inc(refl)}(s,p)$(Eq.(\ref{husimi})), where the origin of $s$ is the
right end of the stadium. 
Figure 2 shows the incident Husimi functions 
$H^{inc}(s,p)$ from which we can realize two
characteristics of the type-I scarred resonances.
First, the critical lines, $ p=\pm p_c$, cross
the localized spots, which explains
why the emission pattern matches with the scarring pattern inside.
This also implies that the type-I scarred resonances are somewhat leaky,
i.e., relatively low $Q$ resonances.
The second is that peak positions $(s^{in\pm}_i,p^{in\pm}_i)$ ($i=1,2,3,4$) 
of the localized spots in $H^{inc}(s,p)$ are not located 
on the expected positions $(s^*_i,\pm p^*_i)$ (black dots in Fig.2) 
from the corresponding periodic ray orbits.
They are slightly shifted in opposite ways depending
on the chirality of propagating component.
Moreover, the direction of the shift appears to be opposite 
in both scarred resonances, e.g., for positive chiral component ($p>0$)
the shift, $\delta s^+_i=s^{in+}_i-s^*_i$, is negative 
for the diamond case and positive for the rectangle case.

\begin{figure}

\vspace{0.3cm}

\hspace{-0.5cm} \includegraphics[height=6cm, width=8cm]{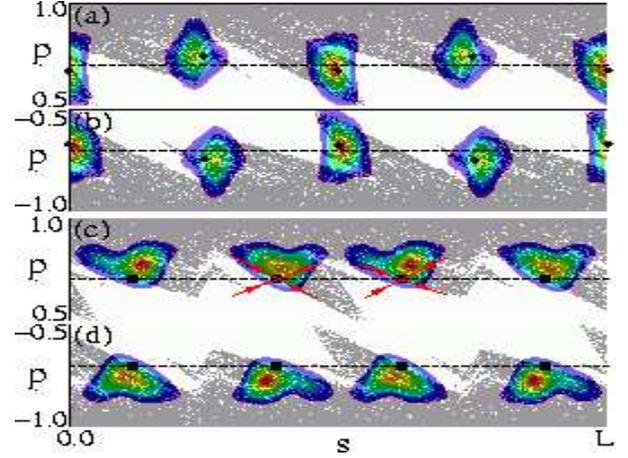}

\vspace{0.0cm}

\caption{(color online) The $H^{inc}(s,p)$ of
the type-I scarred resonances in Fig.1.  (a) and (b) diamond scarring case.
(c) and (d) rectangle scarring case. The SPD is shown by the gray dots and the 
critical line $p_c$ is denoted by the dashed line. 
The bouncing positions $(s^*_i,\pm p^*_i)$ expected from the diamond and rectangle
 ray periodic orbits are pointed as black dots.}

\end{figure}

In the case of a chaotic billiard,  the Husimi function of 
a scarred eigenfunction shows localization on an unstable periodic orbit 
and along nearby stable and unstable manifolds, 
and the resulting hyperbolic localization
reflects that the system has the time-reversal symmetry\cite{Lee03}. 
However, in the chaotic microcavity case the openness 
breaks the time-reversal symmetry and 
makes some structure in the phase space, i.e., the structure of the SPD.
The SPD structure of the present system is shown in the background
in Fig. 2. We note that the expected ray positions $(s^*_i,\pm p^*_i)$ are
on the edge of the structure near the critical line, and the localized
spots of $H^{inc}(s,p)$ are shifted toward the denser part of the
SPD structure. This shift can be understood
from the fact that the rays in the denser part
contribute strongly to the interference process for formation of
a resonance since the denser part of the SPD corresponds to 
the rays surviving longer time.

A more direct explanation is possible in the $H^{inc}(s,p)$
of the rectangle case due to the simplicity of the manifold structure. 
The stable and unstable manifolds emanating from the rectangle periodic orbit
are shown in Fig. 2 (c).
Unlike the hyperbolic localization for a scarred eigenfunction in billiard\cite{Lee03},
the localized spots appear to be deformed showing a part of the 
hyperbolic structure of the manifolds.
This indicates the absence of an inflow along one stable manifold 
in the lower population region in the SPD due to the
refractive escapes from the microcavity. 
As a result, the shifts to denser parts of the SPD not only 
explain the disagreement between  $(s^*_i,\pm p^*_i)$ and $(s^{in\pm}_i,p^{in\pm}_i)$,
but also determine the direction of the shift.

\begin{figure}

\vspace{0.3cm}

\hspace{0.cm}  \includegraphics[height=6cm, width=8.5cm]{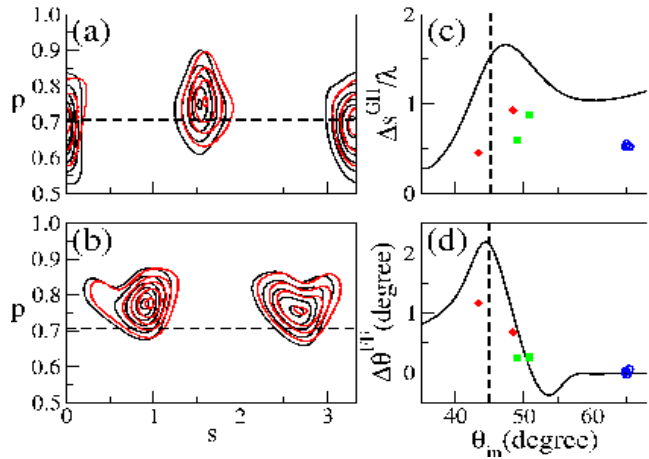}

\vspace{0.0cm}

\caption{ (color online) 
 Contour plots of $H^{inc}(s,p)$(black line) and $H^{refl}(s,p)$(red line)
for the diamond case (a) and the rectangle case (b).  
The Goos-H\"{a}nchen shifts (c) and the angular shifts (d) with the incident
angle  $\theta_{in}=\arcsin |p^{in\pm}|$.  
The solid diamonds and solid rectangles
represent results of the corresponding scarred resonances. The open blue circles are 
for the hexagonally scarred resonance in Fig. 5 (a).
The solid line is the result of a numerical calculation for 
a Gaussian incident beam at planar interface\cite{La86}.
 }

\end{figure}

Note that the incident angles of the scarred patterns are always greater
than the incident angles expected from the corresponding periodic ray orbits,
 i.e., $|p^{in\pm}_i| > p^*_i$. This means that the scarring pattern does not
correspond to a simple connection of $s^{in\pm}_i$.
In order to know the scar beam trajectories, we need to know
positions of localized spots in the reflective Husimi function $H^{refl}(s,p)$. 
From the difference between $H^{inc}(s,p)$ and $H^{refl}(s,p)$,
we can then estimate the Goos-H\"{a}nchen shifts\cite{Go47} and
angular shifts of the reflected beams, i.e.,  
$(\Delta s^{GH\pm}_i,\Delta p^{FF\pm}_i)
=(|s^{out\pm}_i-s^{in\pm}_i|,\pm(p^{out\pm}_i-p^{in\pm}_i))$.
Since the angular shift $\Delta p^{FF}_i$ originates from the partial 
refractive escape from the microcavity, it can be interpreted as
the Fresnel filtering effect on the reflected beam and expected to be
reduced as the refractive escape decreases. 
In Fig. 3 (a) and (b) are shown the differences between $H^{inc}(s,p)$ and
$H^{refl}(s,p)$ for both type-I scarred resonances.
While the Goos-H\"{a}nchen shifts are clearly seen in both cases, 
the angular shift is evident only in the diamond case due to large
refractive escape. 
The $\Delta s^{GH\pm}_i$ and $\Delta \theta^{FF}_i$  are plotted 
as a function of incident angle $\theta_{in}$ in Fig. 3 (c) and (d), 
respectively. The  Goos-H\"{a}nchen shifts distribute around $0.5\lambda$,
$\lambda=2\pi/k$, and the angular shifts decrease with increasing $\theta_{in}$.
This result is qualitatively consistent with the numerical calculation(solid line)
for a Gaussian beam incident on a planar interface\cite{La86}.

\begin{figure}

\vspace{0.3cm}

\hspace{0.cm} \includegraphics[height=5.cm, width=8.2cm]{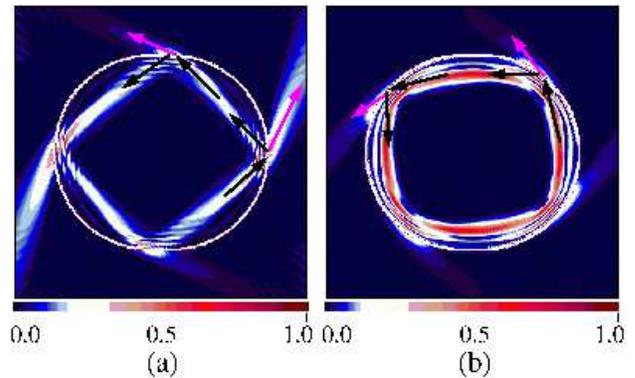}

\vspace{0.0cm}

\caption{(color online) The positive chiral components in the type-I scarred resonance.
 (a) diamond case. (b) rectangle case.  The arrows correspond to the peak positions 
$(s^{in,out+}_i,p^{in,out+}_i)$ of Husimi functions.
Note that the patterns are slightly rotated from the symmetric periodic orbits.
The negative chiral components are just the images of mirror reflection.
} 

\end{figure}

In Fig. 4 the arrows inside the microcavity correspond to
the peak positions $(s^{in,out+}_i,p^{in,out+}_i)$ of both
Husimi functions and the arrows for refractive escapes
are obtained from the replacement of $P_s(s,p)$ by $H^{inc}(s,p)$
in Eq.(\ref{far}).
The background plots are the positive chiral components
obtained by filtering the negative chiral components in the BEM. 
It is clearly shown that the internal patterns are
slightly rotated in opposite directions in both type-I scarred
resonances, and show a good agreement with the arrows.
The same plots for the negative chiral components are
just the mirror images of the positive ones. 
As a result, in the type-I scarred resonances the spatial
splitting of two chiral components are clearly seen.

While the localized spots in $H^{inc}(s,p)$ of type-I scarred resonance
are located near the critical line as shown in Fig. 2, 
in type-II the localized spots are far above the critical line, 
which means that type-II scarred resonances have much higher $Q$ factor
due to strong confinement of light by total internal reflection and
 the refractive emission is very weak.
In Fig. 5 (a) is shown an example of type-II scarred resonance 
showing a hexagonal scarring pattern which is drawn on logarithmic
scale for visibility of the weak refractive emission beams.
From this figure we can find a striking feature of the type-II scarred resonance 
that light emissions do not match with the scarring pattern inside the microcavity. 
In this hexagonal example, although the number of the bouncing points of the scarring
pattern is six, the number of the emitting points is four.
This mismatch between the scarring pattern inside and the emitting pattern
outside is very important in practical experiments 
because the type-II scarred modes cannot be inferred from far field measurements.

\begin{figure}

\vspace{0.3cm}

\hspace{0.cm} \includegraphics[height=8.5cm, width=8.2cm]{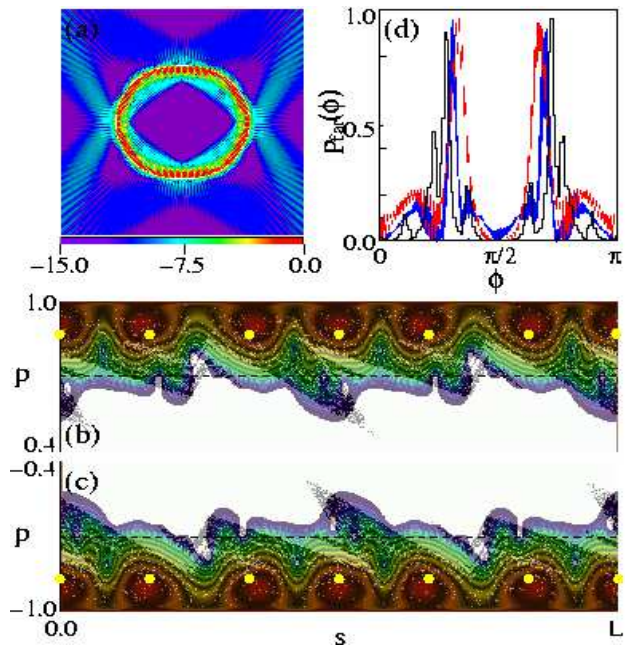}

\vspace{0.0cm}

\caption{(color) (a) Type-II scarred resonance with $kR=(91.17,-0.0137)$ on logarithmic scale. 
(b) and (c) $H^{inc}(s,p)$ on logarithmic scale and
the SPD structure. The yellow dots represent the periodic hexagonal orbit, $(s^*_i,\pm p^*_i)$. 
(d) Far field distributions. The histogram is the result of the SPD  and 
the red line is for the type-II scarred resonance shown in (a), and the blue line is
for another hexagonal scarred resonance with higher $k$, $kR=(170.58,-0.0117)$. }

\end{figure}

Although the $H^{inc}(s,p)$ of the hexagonal resonance 
simply shows, on the normal scale, six spots near the line of $p=\pm 0.90 $,
details near the critical line can be visible in logarithmic scale plot 
shown in Fig. 5 (b) and (c). Here, we emphasize that the structure of $H^{inc}(s,p)$
below the critical line resembles closely that of SPD. 
This implies that a few of rays consisting in the 
hexagonal scarring pattern diffuses chaotically and escapes below 
the critical line and this process happens on the structure
of SPD. Note that the structure of SPD reveals the degree of
the openness on the unstable manifold structure.
Therefore, the SPD structure near the critical line is the major factor
to determine the emission patterns of type-II scarred resonances.

The $P_{far}(\phi)$ based on the SPD structure in Eq.(\ref{far}) is
shown in Fig. 5 (d) as a histogram. 
This shows a good agreement with the far field distribution (red line)
of the type-II resonance shown in Fig. 5 (a). 
We note that the ray dynamical analysis always corresponds to the limit
of $k\rightarrow \infty$. 
The small discrepancy of emission angles would 
be therefore reduced as $k$ increases.
The blue line in Fig. 5 (d) is the $P_{far}(\phi)$ of another
hexagonally scarred resonance with higher $k$, and shows
an improved agreement with the SPD result.

In conclusion, we describe the properties of scarred resonances
in a chaotic microcavity. Their interesting behaviors have been
explained in terms of the SPD in ray dynamics.
The type-I scarred resonances are characterized by the correspondence
between the scarring and the emission patterns and
the spatial splitting of two chiral components. 
On the other hand, the characteristics of type-II scarred
resonances are the mismatch of the emission pattern with the 
scarring pattern inside, and the unique emission pattern given by the SPD.
The results would be very crucial to understand the directionality
and internal patterns of scarred lasing modes generated from a chaotic microcavity
in practical experiments.

We thank T. Harayama, Y.-H. Lee, S.-W. Kim, L. Xu, J. Wiersig for
useful discussions during the NCRICCOC Workshop at Pai-Chai University. 
This work is supported by Creative Research Initiatives of the Korean Ministry of
Science and Technology.


\end{document}